\documentclass[aps,pra,superscriptaddress, 10pt,twocolumn]{revtex4-2}
\usepackage{amsmath}
\usepackage{amssymb}
\usepackage{slashed}
\usepackage{float}
\allowdisplaybreaks
\usepackage{dcolumn}
\newcolumntype{.}{D{x}{}{-1}}
\newcolumntype{d}[1]{D{.}{.}{#1}}
\newcolumntype{w}[1]{D{.}{.}{#1}}
\newcommand*{\centt}[1]{\multicolumn{1}{c}{#1}}

\usepackage{nicefrac}
\usepackage{braket}
\newcommand{\Za}{Z\alpha}
\newcommand{\vare}{\varepsilon}
\newcommand{\lbr}{\langle}
\newcommand{\rbr}{\rangle}

\usepackage{xcolor}

\begin{document}
\title{Second-order hyperfine correction to H, D, and $^3$He energy levels}

\author{Krzysztof Pachucki}
\affiliation{Faculty of Physics, University of Warsaw,
             Pasteura 5, 02-093 Warsaw, Poland}

\author{Vojt\v{e}ch Patk\'o\v{s}}
\affiliation{Faculty of Mathematics and Physics, Charles University,  Ke Karlovu 3, 121 16 Prague
2, Czech Republic}

\author{Vladimir A. Yerokhin}
\affiliation{Max–Planck–Institut f\"ur Kernphysik, Saupfercheckweg 1, 69117 Heidelberg, Germany}

\begin{abstract}
The complete second-order hyperfine-interaction correction is calculated for centroid energy levels of H, D, and $^3$He atoms.
For $^3$He, the corrections of $-2.075$~kHz  and $-0.305$~kHz beyond the leading hyperfine-mixing contribution
are obtained for the  $2^1S$ and $2^3S$ states, respectively.
These results shift the nuclear charge radii difference derived from
the $^3$He\,--$^4$He isotope shift and largely resolve the previously reported disagreement
between the muonic and electronic helium determinations [van der Werf {\em et al.}, arXiv:2306.02333 (2023);
Schuhmann {\em et al.}, arXiv:2305.11679 (2023)].
\end{abstract}
\maketitle

\section{Introduction}

The nuclear magnetic moment causes a shift of atomic energy levels, depending on the total angular momentum
$\vec{F} = \vec{I} + \vec{J}$, where $\vec{I}$ is the nuclear spin and $\vec{J}$ represents the electron angular momentum
\cite{drake:05:springer}. This effect is known as hyperfine splitting (HFS). In hydrogen-like atoms, it is relatively small compared to
the fine structure, which arises from the electron’s spin-orbit coupling.

Due to its small magnitude, hyperfine interaction is typically treated using first-order perturbation theory. The second-order
HFS correction has rarely been addressed in the literature, and when it has, it is often treated incorrectly, primarily due
to the presence of crossed diagrams \cite{pachucki:24:hfs} and the fact that it diverges in the point-nucleus limit.
A proper analysis of the second-order HFS correction has been conducted
only for hydrogen and muonic hydrogen ($\mu$H) atoms so far, in the context of the two-photon exchange contribution \cite{franziska1, franziska2}.
This was required because the hydrogen HFS has been measured precisely \cite{essen:73} and accurate measurements
of HFS in light muonic atoms are underway \cite{crema1, crema2, strasser:23, famu}.

The second-order HFS corrections modify not only the HFS but also the centroid energies. The corresponding
correction has usually been neglected due to its small size, with the leading order
in hydrogenic systems starting at $(Z \alpha)^6 m^3/M^2$, where $M$ is the nuclear mass, $m$ is the electron mass,
$Z$ is the nuclear charge number, and $\alpha$ is the fine-structure constant. This correction forms part of a complete
quadratic recoil correction, which has recently been derived in Ref. \cite{pachucki:24:hpqed}.

In helium-like atoms, the hyperfine interaction causes mixing between triplet and singlet states, which results in significant corrections.
For instance, the HFS of the helium $2^3S$ state is impacted by it by as much as 60 kHz \cite{sternheim, pachucki:01:jpb},
exceeding even the relativistic correction \cite{pachucki:01:jpb}. Similarly, the $2^3S$\,--\,$2^1S$ mixing contributes a substantial correction
to the $^3$He\,--\,$^4$He isotope shift, around 80 kHz \cite{pachucki:15:jpcrd}.

In light of the discrepancy in the nuclear charge radii difference between $^3$He and $^4$He, as determined by electronic
versus muonic measurements \cite{schuhmann:23, werf:23, pachucki:24:rmp},
examining all possible corrections to the $^3$He\,--\,$^4$He isotope shift
has become critically important.
Recently, Qi and collaborators \cite{qi:24} found that the mixing with $n>2$ excited states yields an additional correction to
the helium isotope shift of about 1.37 kHz, which partially explains this discrepancy. In this work, we calculate the complete second-order HFS
correction, finding an even larger correction of 1.77 kHz. Combined with an improved value of the field shift contribution,
this result brings the electronic and muonic
determinations of the nuclear charge radii difference \cite{schuhmann:23, werf:23} into a
$1.3\,\sigma$ agreement.

\section{Hydrogenic atom}
Let us begin by considering the HFS interaction in hydrogenic systems.
The Dirac Hamiltonian in the external electromagnetic field  is (in natural units $\hbar = c = 1$)
\begin{align}
H_\mathrm{EM} =&\ \vec\alpha\cdot(\vec p-e\,\vec A_\mathrm{hfs}) + \beta\,m + e\,A^0
\nonumber \\ =&\
H_\mathrm{D} + V_\mathrm{hfs}, \label{01}
\end{align}
where $e$ is the electron charge, $A_0$ is the Coulomb field from the nucleus
\begin{align}
V \equiv e\,A^0 = -\frac{Z\,\alpha}{r}\,, \label{02}
\end{align}
and $\vec A_\mathrm{hfs}$ is the vector potential from the nuclear magnetic moment $\vec \mu$
\begin{align}
\vec A_\mathrm{hfs} = \frac{1}{4\,\pi}\,\vec \mu\times\frac{\vec r}{r^3}\,. \label{03}
\end{align}
The corresponding hyperfine interaction for a point nucleus is
\begin{align}
V_\mathrm{hfs} =&\  -e\,\vec\alpha\cdot\vec A_\mathrm{hfs}
\nonumber \\ =&\
\frac{e}{4\,\pi}\,\vec \mu\cdot\vec\alpha\times\frac{\vec r}{r^3}\,. \label{04}
\end{align}
The second-order hyperfine correction to energy is uv divergent for a point nucleus.
Therefore, we cannot ignore the charge and magnetic moment nuclear form factors,
but for simplicity we assume that they are equal to each other, thus
\begin{align}
\rho_E(r) = \rho_M(r) = \rho(r)\,, \label{05}
\end{align}
and neglect the higher order nuclear electromagnetic moments. Thus, potentials take the general form
\begin{align}
V =&\ -Z\,\alpha\,\biggl[\frac{1}{r}\biggr]_\mathrm{fs}\equiv -\frac{\Za}{r}\,F_C(r)\,,\label{06} \\
V_\mathrm{hfs} =&\ \frac{e}{4\,\pi}\,\vec \mu\cdot\vec\alpha\times\biggl[\frac{\vec r}{r^3}\biggr]_\mathrm{fs}
\equiv \frac{e}{4\,\pi}\,\vec \mu\cdot\,\frac{\vec\alpha\times\vec r}{r^3} \,F_{\rm hfs}(r)
\,, \label{07}
\end{align}
and $F_C(r) = F_{\rm hfs}(r) = 1$ for the point nucleus.

The second-order hyperfine correction to energy in the hydrogenic system  is
\begin{align}
E_{\rm hfs2} =&\ \langle \phi| V_\mathrm{hfs}\,\frac{1}{(E_D-H_D)'}\, V_\mathrm{hfs} |\phi\rangle\,. \label{08}
\end{align}
The product of the nuclear magnetic moments in the above is to be replaced by
$ \mu^a\,\mu^b \rightarrow  \delta^{ab}\,\vec\mu^{\,2}/3$, because here we are interested
in corrections to energy levels but not to the hyperfine structure.
Using the spectral representation of the Green function, we express Eq.~(\ref{08}) as
\begin{align}
E_{\rm hfs2} =&\ \frac13\,(\Za)^2\,m^3\,\left(\frac{g}{2M}\right)^2\,I(I+1)\,
\sum_{n,\kappa_n,\mu_n}\!\!\!^{'}\,
 \frac1{\vare_a-\vare_n}
 \nonumber \\ &
 \times
  \lbr n_a\kappa_a\mu_a| F_{\rm hfs}(r)\,\frac{(\vec{\alpha} \times \vec{r})_i}{r^3}|n\kappa_n\mu_n\rbr\,
 \nonumber \\ &
 \times
  \lbr n\kappa_n\mu_n| F_{\rm hfs}(r)\,\frac{(\vec{\alpha} \times \vec{r})_i}{r^3}|n_a\kappa_a\mu_a\rbr\,, \label{10}
\end{align}
where we introduced the nuclear $g$-factor $g$ as
\begin{align}
\vec\mu = \frac{q}{2\,M}\,g\,\vec I, \label{09}
\end{align}
where $q= Z\,|e|$.
Performing the angular-momentum algebra in Eq.~(\ref{10}), we obtain, for the $1s$ reference state
($\kappa_a = -1$),
\begin{align}
E_{\rm hfs2} = \frac23\,(\Za)^2\,m^3\,\left(\frac{g}{2M}\right)^2\,I(I+1)\, {\cal E}\,, \label{12}
\end{align}
where ${\cal E}$ is
\begin{align}
{\cal E} =&\ \sum_{\kappa_n = -1,2}\,
 a_{\kappa_n}\,
\sum_{n}\,\!^{'}\,
 \frac1{\vare_a-\vare_n}
 \nonumber \\ &\
 \times
    \left\{
    \int_0^{\infty}x^2dx\,\frac{F_\mathrm{hfs}(x)}{x^2}\,
    [g_a(x)\,f_n(x)+f_a(x)\,g_n(x)]\right\}^2
    \,. \label{13}
\end{align}
Here, $a_{-1} = 2/3$, $a_2 = 1/3$, and $g_a(x)$ and $f_a(x)$ are the upper and the lower radial
components of the Dirac wave function of the reference state.
For the dipole parametrization with  $r_M = r_C$, the
magnetic and electric distribution functions are
\begin{align}
F_{\rm hfs}(x) =&\ 1 - e^{-\lambda x}
 \Big[ 1 + \lambda x + \frac12 (\lambda x)^2\Big]\,, \nonumber \\
F_{C}(x) =&\  1 - \Big(1 + \frac{\lambda x}{2} \Big) \,e^{-\lambda x} \,, \label{15}
\end{align}
with $\lambda = 2\sqrt{3}/r_C$.
We evaluate numerically Eq.~(\ref{13}) for the dipole nuclear model by representing the
summation over the spectrum as a sum over the pseudostates
in the basis of B-splines \cite{johnson:88,shabaev:04:DKB}.
A series of calculations were performed for a fixed nuclear charge $Z$ but for different
nuclear radii $r_C$. The obtained results were fitted to the form
\begin{align}
{\cal E}(r_C) = (\Za)^4\,\big[ a\ln r_C + b + c\, r_C + \ldots \big]\,, \label{16}
\end{align}
which yielded the following result for the leading logarithmic coefficient,
\begin{align}
a = 5.00\,(1)\,,  \label{17}
\end{align}
in agreement with the result of  analytical calculations in Eq. (\ref{76}).
Next, we performed numerical calculations of ${\cal E}$ for different values of $Z$ with a
fixed value of $r_C$ and fitted the results to the form
\begin{align}
{\cal E}(\Za) = (\Za)^4\,\big[ 5\ln \left(\Za\,m\,r_C\right) + B + \Za\,C+ \ldots \big]\,. \label{18}
\end{align}
We obtained the nonlogarithmic coefficient $B$ as
\begin{align}
B = 2.700\,(2)\,, \label{19}
\end{align}
again in agreement with the analytical  result  in Eq. (\ref{76}).
The numerical results for the leading $\Za$-expansion coefficients
serve as a valuable test of the correctness of the analytical derivation for few-electron atoms.

\section{NRQED approach for a few electron systems}
For light few-electron systems, the best computational approach is that based on nonrelativistic quantum electrodynamics (NRQED),
because we can account accurately for relativistic effects and electron correlations on the same footing.
In fact, the derivation is similar to that of the internuclear spin-spin interaction in molecules by Ramsey \cite{ramsey:53},
with the difference that here we have the same nucleus and thus have to include the hard electron-nucleus interaction,
which is highly singular.
Therefore, we will use the dimension regularization for the derivation of NRQED operators, see Appendix~\ref{app:dimreg} for details.
Because we know that the shift is proportional to $\vec I^{\,2}$, we can assume initially that $I=1/2$
and later after cancellation of $1/\varepsilon$ terms, replace it by an arbitrary $I$.
Therefore, in $d=3-2\,\varepsilon$ dimensions the vector potential takes the form
\begin{align}
A_\mathrm{hfs}^i(\vec q) =&\ -i\,Z\,e\,\frac{g}{4\,M}\,\sigma_N^{ij}\,\frac{q^j}{q^{2}}\,\tilde\rho(q^{2})\,, \label{20}
\end{align}
where $\tilde\rho$ is a Fourier transform of $\rho$ in Eq. (\ref{05}) and $\sigma_N^{ij}$ is $d$-dimensional spin operator for the nucleus defined in Eq. (\ref{A01}).
In the coordinate representation it is
\begin{align}
A_\mathrm{hfs}^i(\vec r) =&\ -i\,Z\,e\,\frac{g}{4\,M}\,\sigma_N^{ij}\,\int\frac{d^d q}{(2\,\pi)^d}\,e^{i\,\vec q\cdot\vec r}\,\frac{q^j}{q^{2}}\,\tilde\rho(q^{2})
\nonumber \\ =&\
\frac{Z\,e}{4\,\pi}\,\frac{g}{4\,M}\,\sigma_N^{ij}\,\biggl[\frac{r^j}{r^3}\biggr]_\mathrm{fs, \varepsilon}
.\label{21}
\end{align}
In $d=3$ dimensions, $\sigma^{ij} = \epsilon^{ijk}\,\sigma^k$ and for a point nucleus $A_\mathrm{hfs}^i(\vec r)$ becomes
\begin{align}
\vec A_\mathrm{hfs}(\vec r) =&\ -\frac{Z\,e}{4\,\pi}\,\frac{g}{2\,M}\,\vec I\times\frac{\vec r}{r^3}
=
\frac{1}{4\,\pi}\,\vec\mu\times\frac{\vec r}{r^3}
\,,\label{22}
\end{align}
as it should.

The leading second-order hyperfine-interaction correction is of the order $m\,\alpha^6\,(m/M)^2$.
We represent it as a sum of three terms
\begin{align}
E_\mathrm{hfs2} =&\ E_\mathrm{sec} + E_\mathrm{diam} + E_\mathrm{high}\,,  \label{23}
\end{align}
which we calculate in the following three sections.
Similarly to the first-order HFS correction, which is proportional to the Fermi coefficient $E_F$,
the second-order HFS correction is proportional to the
second-order Fermi coefficient  $E_{F2}$,
\begin{align}
E_{F2} =
\alpha^6\,m\,\Bigl(\frac{Z\,g\,m}{2\,M}\Big)^2\,I(I+1) \,. \label{EF2}
\end{align}

\section{$E_\mathrm{sec}$}
$E_\mathrm{sec}$ is a second-order correction of the form
\begin{align}
E_\mathrm{sec} =&\ \bigg \langle H_\mathrm{hfs}\,\frac{1}{E-H}\,H_\mathrm{hfs} \bigg \rangle\,,  \label{24}
\end{align}
where $H$ is the nonrelativistic Hamiltonian
\begin{align}
H = \frac{p_1^2}{2\,m} + \frac{p_2^2}{2\,m} + \biggl[\frac{\alpha}{r_{12}} -\frac{Z\,\alpha}{r_1} - \frac{Z\,\alpha}{r_2}\biggr]_\varepsilon,  \label{25}
\end{align}
and $H_\mathrm{hfs}$ is the leading hyperfine interaction.
It is obtained from the nonrelativistic expansion of $V_\mathrm{hfs}$ for a point nucleus
\begin{align}
V_\mathrm{hfs} =&\ -\frac{Z\,\alpha\,g}{4\,M}\,\alpha^i\,\sigma_N^{ij}\,\biggl[\frac{r^j}{r^3}\biggr]_\mathrm{\varepsilon}\,,  \label{26}
\end{align}
namely
\begin{align}
H_\mathrm{hfs} =&\
-\frac{Z\,\alpha\,g}{4\,M}\,\sigma_N^{ij}\,\biggl\{\sigma^i\,\biggl[\frac{r^j}{r^3}\biggr]_\mathrm{\varepsilon}\,,\,\frac{\sigma^k\,p^k}{2\,m}\biggr\}\,,  \label{27}
\end{align}
where $\{A\,,\,B\} = A\,B + B\,A$. In $d$-dimensions
\begin{align}
\sigma^i\,\sigma^k =&\ \frac{1}{2}\,\{\sigma^i\,,\,\sigma^k\} + \frac{1}{2}\,[\sigma^i\,,\,\sigma^k]
\nonumber \\ =&\
\delta^{ik} + i\,\sigma^{ik}\,,  \label{28}
\end{align}
thus
\begin{align}
H_\mathrm{hfs} =&\
-\frac{Z\,\alpha\,g}{4\,m\,M}\,\sigma_N^{ij}\,\biggl(\sigma^{ik}\,\frac{i}{2}\,\biggl[\frac{r^j}{r^3}\,,\,p^k\biggr]_\mathrm{\varepsilon}
+\biggl[\frac{r^j}{r^3}\biggr]_\mathrm{\varepsilon} p^i\biggr)
\nonumber \\ =&\
H_\mathrm{hfsA} + H_\mathrm{hfsB} + H_\mathrm{hfsC}\,,  \label{29}
\end{align}
where after summing over all electrons
\begin{align}
H_\mathrm{hfsA} =&\
\sum_a \frac{Z\,\alpha\,g}{2\,m\,M}\,\sigma_N^{ij}\,\sigma_a^{ij}\,\frac{\pi}{d}\,\delta^d(r_a)\,,  \label{30}
\\
H_\mathrm{hfsB} =&\
\sum_a \frac{Z\,\alpha\,g}{8\,m\,M}\,\sigma_N^{ij}\,\sigma_a^{ik}\,\biggl[\frac{\delta^{jk}}{r_a^3} - 3\,\frac{r_a^j\,r_a^k}{r_a^5}\biggr]_\mathrm{\varepsilon}\,, \label{31}
\\
H_\mathrm{hfsC}
\stackrel{d=3}{=}&\ \sum_a \frac{Z\,\alpha\,g}{2\,m\,M}\,\vec I\cdot\frac{\vec r_a}{r_a^3}\times \vec p_a\,. \label{32}
\end{align}
The corresponding second-order contributions
\begin{align}
E_\mathrm{sec} =&\  E_\mathrm{secA} + E_\mathrm{secB} + E_\mathrm{secC} \label{33}
\end{align}
are angular averaged [see Eqs. (\ref{A02}) and (\ref{A03}) in Appendix~\ref{app:dimreg}] and expressed in  atomic units.

We proceed first with the calculation of $E_\mathrm{secA}$,
\begin{align}
E_\mathrm{secA} =&\ \bigg \langle H_\mathrm{hfsA}\,\frac{1}{(E-H)'}\,H_\mathrm{hfsA} \bigg \rangle
\nonumber \\ =&\
\frac{E_{F2}}{d^3\,(d-1)}\,\sum_{a,b}
\nonumber \\ \times&\
\bigg \langle \sigma_a^{ij}\,4\,\pi\,\delta^d(r_a)\,\frac{1}{(E-H)'}\,\sigma_b^{ij}\,4\,\pi\,\delta^d(r_b) \bigg \rangle\,, \label{34}
\end{align}
where atomic units are used for the matrix elements, and $E_{F2}$ is assumed to be the $d$-dimensional generalization of Eq.~(\ref{EF2}) with $I(I+1)\to \langle \vec{I}^{\,2}\rbr_\varepsilon$.
The expression in Eq.~(\ref{34}) is singular and needs to be transformed to the regular form
by pulling out the $1/\varepsilon$ singularity from the second-order matrix elements.
This is achieved by the use of the identity
\begin{align}
H_a|\phi\rangle \equiv&\ \{H-E\,,\,Q_a\}|\phi\rangle + V_{a}|\phi\rangle\,, \label{36}
\end{align}
where $\phi$ is the reference state,
\begin{align}
H_a =&\ 4\,\pi\,\delta^d(r_a)\,, \label{37}\\
Q_a =&\ 2\,\biggl[\frac{1}{r_a}\biggr]_\varepsilon\,, \label{38} \\
V_{a} =&\ -2\,\biggl[\frac{\vec r_a}{r_a^3}\biggr]_\varepsilon\cdot\vec\nabla_a\,, \label{39}
\end{align}
to obtain
\begin{align}
	\bigg\langle H_a\,\frac{1}{(E-H)'}\,H_b\bigg\rangle = &\
	\bigg\langle V^+_a\,\frac{1}{(E-H)'}\,V_b\bigg\rangle
	+ \langle H_a \rangle\,\langle  Q_b\rangle
 \nonumber \\
	 &\ + \langle H_b \rangle\,\langle  Q_a\rangle	
	-\,\langle H_a\,Q_b\rangle
	-\,\langle H_b\,Q_a\rangle \nonumber \\
	 &\ + \langle Q_a\,(H-E)\, Q_b\rangle
\,, \label{40}
\end{align}
where $V^+_a$ is a Hermitian conjugation of $V_a$.
The resulting form of $E_\mathrm{secA}$ is
\begin{widetext}
\begin{align}
E_\mathrm{secA} =&\  E_{F2}\,\frac{1}{d^3\,(d-1)}\,
\sum_{a,b}\sigma_a^{ij}\, \sigma_b^{ij}\,
	\biggl[\bigg\langle V^+_a\,\frac{1}{(E-H)'}\,V_b\bigg\rangle
	+ \langle 4\,\pi\,\delta^3(r_a) \rangle\,\bigg\langle  \frac{2}{r_b}\bigg\rangle
	+ \langle 4\,\pi\,\delta^3(r_b) \rangle\,\bigg\langle  \frac{2}{r_a}\bigg\rangle
 \nonumber \\ &\	
	-\,\bigg\langle 4\,\pi\,\delta^d(r_a)\,2\,\biggl[\frac{1}{r_b}\biggr]_\varepsilon\bigg\rangle
	-\,\bigg\langle 4\,\pi\,\delta^d(r_b)\,2\,\biggl[\frac{1}{r_a}\biggr]_\varepsilon\bigg\rangle
	+ 2\,\delta_{ab}\,\bigg\langle \biggl[\frac{\vec r_a}{r_a^3}\biggr]_\varepsilon\cdot\biggl[\frac{\vec r_b}{r_b^3}\biggr]_\varepsilon\bigg\rangle\biggr]
\,. \label{41}
\end{align}	

For the second-order tensor contribution $E_\mathrm{secB}$ we use Eq. (\ref{A03}) to contract indices, and
\begin{align}
E_\mathrm{secB} =&\ \bigg \langle H_\mathrm{hfsB}\,\frac{1}{E-H}\,H_\mathrm{hfsB} \bigg \rangle
= E_{F2}\,\frac{(d-2)}{2\,d^2\,(d-1)^2}\,\sum_{a,b}\,
\bigg \langle \sigma_a^{kl}\,\biggl[\frac{\delta^{ij}}{r_a^3} - 3\,\frac{r_a^i\,r_a^j}{r_a^5}\biggr]_\mathrm{\varepsilon}\,\frac{1}{E-H}\,
\sigma_b^{kl}\,\biggl[\frac{\delta^{ij}}{r_b^3} - 3\,\frac{r_b^i\,r_b^j}{r_b^5}\biggr]_\mathrm{\varepsilon} \bigg \rangle\,, \label{42}
\end{align}
 we use the following identity to pull out $1/\varepsilon$ singularity from the matrix elements:
\begin{align}
	H_a^{ij}|\phi\rangle  \ \equiv&\ \{H-E, Q_a^{ij} \}|\phi\rangle + V_a^{ij}|\phi\rangle\,, \label{43}
\end{align}
where
\begin{align}	
        H_a^{ij} =&\ \bigg[\frac{\delta^{ij}}{r_a^3}-3\frac{r_a^i r_a^j}{r_a^5}\bigg]_\varepsilon
                     = \Big(-\nabla^i\,\nabla^j + \frac{\delta^{ij}}{d}\,\vec\nabla^{\,2}\Big)\,\mathcal{V}(r_a)\,, \label{44}\\
	Q_a^{ij} =&\ 2\,\biggl[\frac{\delta^{ij}}{6\,r_a} - \frac{r_a^i\,r_a^j}{2\,r_a^3}\biggr]_\mathrm{\varepsilon}
	              = 2\,\Big(-\nabla^i\,\nabla^j + \frac{\delta^{ij}}{d}\,\vec\nabla^{\,2}\Big)\,\mathcal{V}^{(2)}(r_a)\,, \label{45} \\
	V_a^{ij}|\phi\rangle =&\ \bigg[-\frac{\delta^{ij} r_a^k + 3 \,\delta^{ik} r_a^j + 3\, \delta^{jk} r_a^i}{3\,r_a^3} + 3 \frac{r_a^i r_a^j r_a^k}{r_a^5}\bigg]\,\nabla_a^k|\phi\rangle\,,
	\label{46}
\end{align}
with $\mathcal{V}$ and $\mathcal{V}^{(2)}$ defined in Appendix~\ref{app:dimreg}. We obtain
\begin{align}
\bigg \langle H_a^{ij}\,\frac{1}{E-H}\, H_b^{ij} \bigg \rangle = &\
	\bigg\langle V_a^{+ij}\,\frac{1}{E-H}\,V_b^{ij}\bigg\rangle
	-\,\langle H_a^{ij}\,Q_b^{ij}\rangle
	-\,\langle H_b^{ij}\,Q_a^{ij}\rangle
	+ \langle Q_a^{ij}\,(H-E)\, Q_b^{ij}\rangle\,. \label{47}
\end{align}
The result for $E_\mathrm{secB}$ is then
\begin{align}
E_\mathrm{secB} =
&\  E_{F2}\,\frac{(d-2)}{2\,d^2\,(d-1)^2}\,\sum_{a,b}\,\sigma_a^{st}\,\sigma_b^{st}\,
\biggl[\bigg\langle V_a^{+ij}\,\frac{1}{E-H}\,V_b^{ij}\bigg\rangle
	-2\,\bigg\langle \bigg[\frac{\delta^{ij}}{r_a^3}-3\frac{r_a^i r_a^j}{r_a^5}\bigg]_\varepsilon\,
	\biggl[\frac{\delta^{ij}}{6\,r_b} - \frac{r_b^i\,r_b^j}{2\,r_b^3}\biggr]_\mathrm{\varepsilon}\bigg\rangle
\nonumber \\ &\	
	-2\,\bigg\langle \bigg[\frac{\delta^{ij}}{r_b^3}-3\frac{r_b^i r_b^j}{r_b^5}\bigg]_\varepsilon\,
	\biggl[\frac{\delta^{ij}}{6\,r_a} - \frac{r_a^i\,r_a^j}{2\,r_a^3}\biggr]_\mathrm{\varepsilon}\bigg\rangle
	+ 2\,\delta_{ab}\,\bigg\langle \nabla_a^k\,\biggl[\frac{\delta^{ij}}{6\,r_a} - \frac{r_a^i\,r_a^j}{2\,r_a^3}\biggr]_\mathrm{\varepsilon}\,
	\nabla_b^k \biggl[\frac{\delta^{ij}}{6\,r_b} - \frac{r_b^i\,r_b^j}{2\,r_b^3}\biggr]_\mathrm{\varepsilon}\bigg\rangle\biggr] \,. \label{48}
	\end{align}
\end{widetext}
Spin-orbit contribution $E_\mathrm{secC}$ does not require any regularization,
\begin{align}
E_\mathrm{secC} =&\ \bigg \langle H_\mathrm{hfsC}\,\frac{1}{E-H}\,H_\mathrm{hfsC} \bigg \rangle
\nonumber \\  =&\
 E_{F2}\, \frac{1}{3}\,\sum_{a,b}
\bigg \langle \frac{\vec r_a}{r_a^3}\times \vec p_a\,\frac{1}{E-H}\,\frac{\vec r_b}{r_b^3}\times \vec p_b \bigg \rangle\,, \label{49}
\end{align}
 and can be calculated directly in $d=3$ dimensions.

\section{ $E_\mathrm{diam}$}
The diamagnetic contribution in natural units is
\begin{align}
 E_\mathrm{diam} = \sum_a \frac{e^2}{2\,m}\,\bigl\langle\vec{\mathcal{A}}(\vec r_{a})^2\bigr\rangle\,, \label{50}
\end{align}
where
\begin{align}
e\,\vec{\mathcal{A}}(\vec r_{a}) = \frac{Z\,\alpha\,g}{4\,M}\,\sigma_N^{ij}\,\biggl[\frac{r_a^j}{r_a^3}\biggr]_\mathrm{\varepsilon}\,.  \label{51}
\end{align}
After transforming to atomic units and performing the angular average, it becomes
\begin{align}
 E_\mathrm{diam} =&\
E_{F2}\,\frac{1}{d}\,\sum_a\biggl \langle\frac{1}{r_a^4}\biggr\rangle_\mathrm{\varepsilon}
\,. \label{52}
 \end{align}

\section{ $E_\mathrm{high}$}
The high-energy contribution is obtained from the scattering amplitude, and is essentially the same as for the hydrogenic case.
In momentum space and natural units, we have
\begin{align}
V(\vec q) =&\ -\frac{4\,\pi\,Z\,\alpha}{\vec q^{\,2}}\,\tilde\rho(q^{2})\,, \label{53}\\
V_\mathrm{hfs}(\vec q) =&\ i\,4\,\pi\,Z\,\alpha\,\frac{g}{4\,M}\,\alpha^i\,\sigma_N^{ij}\,\frac{q^j}{q^{2}}\,\tilde\rho(q^{2})\,.  \label{54}
\end{align}
Then,
\begin{align}
 E_\mathrm{high} = &\
\phi^2(0)\,\int \frac{d^dq_1}{(2\,\pi)^d}\, \int \frac{d^dq_2}{(2\,\pi)^d}\,V(\vec q_1-\vec q_2)
\nonumber \\  \times &\
\langle t|
\gamma^0\,S_F(q_2-q_1)\,\gamma^0\,V_\mathrm{hfs}(-\vec q_1)\,S_F(q_2)\, \gamma^0\,V_\mathrm{hfs}(\vec q_2)
\nonumber \\ + &\
\gamma^0\,V_\mathrm{hfs}(-\vec q_1)\,S_F(q_1)\,\gamma^0\,S_F(q_2)\, \gamma^0\,V_\mathrm{hfs}(\vec q_2)
\nonumber \\ + &\
\gamma^0\,V_\mathrm{hfs}(-\vec q_1)\,S_F(q_1)\,\gamma^0\,V_\mathrm{hfs}(\vec q_2)\,S_F(q_1-q_2)\, \gamma^0 |t\rangle  \label{55}
\,,
\end{align}
where $S_F$ is the fermion propagator and $t = (m,\vec 0)$. Performing traces with Dirac matrices, one obtains

\begin{align}
E_\mathrm{high}=&\
-\phi^2(0)\,(4\,\pi\,Z\,\alpha)^3\,\biggl(\frac{g}{2\,M}\biggr)^2\,\frac{4}{d\,(d-1)}\,\lbr\vec{I}^{\,2}\rbr_\varepsilon\,
\nonumber \\ &\ \times
\int \frac{d^dq_1}{(2\,\pi)^d}\, \int \frac{d^dq_2}{(2\,\pi)^d}\,\frac{\tilde\rho(q_1^2)\,\tilde\rho(q_2^2)\,\tilde\rho(q_3^2)}{q_1^2\,q_2^2\,q_3^2}
\nonumber \\ &\
\times \bigg( 1 + (d-2)\,\frac{(\vec q_1 \cdot \vec q_2)\,(\vec q_2 \cdot \vec q_3)\,(\vec q_3 \cdot \vec q_1)}{q_1^2\,q_2^2\,q_3^2} \bigg)\,,  \label{56}
\end{align}
where $\vec q_3 = \vec q_1-\vec q_2$. After transforming integral to the coordinate representation, it becomes
\begin{widetext}
\begin{align}
E_\mathrm{high}=&\
-\phi^2(0)\,(Z\,\alpha)^3\,\biggl(\frac{g}{4\,M}\biggr)^2\,\frac{16\,\lbr\vec{I}^{\,2}\rbr_\varepsilon}{d\,(d-1)}
\int d^dr
\bigl[ \mathcal{V}_\rho(r)^3 - (d-2)\,\nabla^i \nabla^j \mathcal{V}_\rho^{(2)}(r)
\,  \nabla^j \nabla^k \mathcal{V}_\rho^{(2)}(r)  \, \nabla^k \nabla^i \mathcal{V}_\rho^{(2)}(r) \big]
\nonumber \\ =&\ E_\mathrm{highA} + E_\mathrm{highB}\,,  \label{57}
\end{align}
where $\mathcal{V}_\rho(r)$ and $\mathcal{V}_\rho^{(2)}(r)$ are defined in Appendix~\ref{app:dimreg}.
$E_\mathrm{highA}$ includes the  $\vec r$ integral over a small region of the nuclear charge distribution,
where we can set $d=3$. It is evaluated as
\begin{align}
E_\mathrm{highA} =&\
-\phi^2(0)\,(Z\,\alpha)^3\,\biggl(\frac{g}{4\,M}\biggr)^2\,\frac{8}{3}\,\lbr\vec{I}^{\,2}\rbr
\int^R d^3r \bigl[ {V}_\rho(r)^3 - \nabla^i \nabla^j {V}_\rho^{(2)}(r)\, \nabla^j \nabla^k {V}_\rho^{(2)}(r)\, \nabla^k \nabla^i {V}_\rho^{(2)}(r) \big]\,.  \label{58}
\end{align}
It is convenient at this point to use the explicit form of the nuclear charge distribution,
which we chose to be dipole parametrization, the same as used in the hydrogenic case in Sec. II,
\begin{align}
\tilde\rho(p^2) =&\ \frac{\lambda^4}{(\lambda^2+p^2)^2}\,.  \label{59}
\end{align}
Thus, $d=3$ potentials become
\begin{align}
4\,\pi\,\rho(r) =&\ 4\pi \int \frac{d^3p}{(2\pi)^3}\tilde\rho(p^2)\, e^{i \vec{p}\cdot\vec{r}} =\frac{\lambda^3}{2}\,e^{-\lambda\,r}\,,  \label{60}\\
{V}_\rho(r) =&\ 4\pi \int \frac{d^3p}{(2\pi)^3} \frac{\tilde\rho(p^2)}{p^2} e^{i \vec{p}\cdot\vec{r}} =
 \frac{1}{r} - \frac{e^{-\lambda\,r}}{r} - \frac{\lambda}{2}\,e^{-\lambda\,r}\,,  \label{61}\\
{V}_\rho^{(2)}(r) =&\ 4\pi \int \frac{d^3p}{(2\pi)^3} \frac{\tilde\rho(p^2)\,e^{i \vec{p}\cdot\vec{r}}-1}{p^4} =
-\frac{r}{2} - \frac{2}{\lambda^2\,r} + \frac{1}{2\,\lambda}\,e^{-\lambda\,r} + \frac{2}{\lambda^2}\,\frac{e^{-\lambda\,r}}{r}\,.  \label{62}
\end{align}
Now the $r$-integration in Eq. (\ref{58}) can be performed analytically, with the result
\begin{align}
E_\mathrm{highA} =&\
-\phi^2(0)\,(4\,\pi\,Z\,\alpha)^3\,\biggl(\frac{g}{4\,M}\biggr)^2\,\frac{8}{3}\,\lbr\vec{I}^{\,2}\rbr\,
5\,\pi\,\biggl(-\frac{467}{540} + \frac{2\,\ln 2}{5} + \frac{3\,\ln 3}{10} - \ln \frac{r_C}{R}  +\gamma \biggr)\,,  \label{63}
 \end{align}
where $r_C = 2\,\sqrt{3}/\lambda$.  It should be mentioned that the description of the nucleus through the elastic form-factors $\rho_C$ and  $\rho_M$ here 
is not fully justified. It is because for high momenta $\sim \nicefrac{\hbar}{r_C}$ the electron sees individual nucleons, rather than the nucleus as a whole. 
The detailed analysis of this contribution is complicated and should be performed together with other three-photon exchange corrections  \cite{pachucki:18}. 
We will use the above form-factor result for comparison with direct numerical calculations for hydrogenic systems described in Sec. II, 
keeping in mind that for the $2^3S - 2^1S$ transition in $^3$He it is negligible.

The part $E_\mathrm{highB}$ of the high energy contribution involves integral outside of the nuclear region, thus we can assume the
point nucleus and obtain
\begin{align}
E_\mathrm{highB} =&\
-\phi^2(0)\,(Z\,\alpha)^3\,\biggl(\frac{g}{4\,M}\biggr)^2\,\frac{16}{d\,(d-1)}\,\lbr\vec{I}^{\,2}\rbr_\varepsilon
\int_R d^dr \bigl[ \mathcal{V}(r)^3 - (d-2)\,\nabla^i \nabla^j \mathcal{V}^{(2)}(r)\,\nabla^j \nabla^k \mathcal{V}^{(2)}(r)  \, \nabla^k \nabla^i \mathcal{V}^{(2)}(r) \big]
\nonumber \\ =&\
\phi^2(0)\,(Z\,\alpha)^3\,\biggl(\frac{g}{4\,M}\biggr)^2\,\lbr\vec{I}^{\,2}\rbr_\varepsilon\,
[(4\,\pi)^\varepsilon \Gamma(1+\varepsilon)]^2\, \frac{40}{3}\,\pi\,\biggl(\frac{23}{30} + \frac{1}{4\,\varepsilon} + \ln R + \gamma \biggr)\,,  \label{64}
\end{align}
and by convention the factor $[(4\,\pi)^\varepsilon \Gamma(1+\varepsilon)]^2$ is consistently pulled out from all the terms.

The sum of both parts is independent on arbitrary cutoff $R$, and is
\begin{align}
E_\mathrm{high} =&\ (Z\,\alpha)^3\,\biggl(\frac{g}{2\,M}\biggr)^2\,\lbr\vec{I}^{\,2}\rbr_\varepsilon \frac{10}{3}\,\pi
\biggl(\frac{881}{540} - \frac{2\,\ln 2}{5}
- \frac{3\,\ln 3}{10}
+ \frac{1}{4\,\varepsilon} + \ln r_C  \biggr)\,\langle\delta^d(r)\rangle\,.  \label{65}
\end{align}
Transforming to atomic units by $\vec r \rightarrow (m\,\alpha)^{-1/(1+2\,\varepsilon)}\,\vec r$, one pulls out the factor
$m^{(1-2\,\varepsilon)/(1+2\varepsilon)}\,\alpha^{2/(1+2\,\varepsilon)}$
from $H$ and obtains the nonrelativistic Hamiltonian in atomic units.
Similarly, for $H^{(6)}$ the common factor in atomic units
\begin{equation}
\eta = m^{(1-10\,\varepsilon)/(1+2\varepsilon)}\,\alpha^{6/(1+2\,\varepsilon)}  \label{66}
\end{equation}
is pulled out from all the terms, which corresponds to the replacement
$m\rightarrow 1, \alpha \rightarrow 1$, and
\begin{align}
E_\mathrm{high} = &\
E_{F2}\, \frac{10}{3}\,\pi
\biggl(\frac{881}{540} - \frac{2\,\ln 2}{5} - \frac{3\,\ln 3}{10}
 + \frac{1}{4\,\varepsilon} + \ln m\,\alpha\,r_C  \biggr)\,\sum_a\,Z\,\langle\delta^d(r_a)\rangle\,.  \label{67}
\end{align}

\section{Complete result}
In order to eliminate $1/\varepsilon$ singularity, we transform all singular operators (in atomic units) to the same form,
\begin{align}
	\bigg\langle \bigg[\frac{\delta^{ij}}{r^3}-3\frac{r^i r^j}{r^5}\bigg]_\varepsilon\,
	\biggl[\frac{\delta^{ij}}{6\,r} - \frac{r^i\,r^j}{2\,r^3}\biggr]_\mathrm{\varepsilon}\bigg\rangle =&\ (1-\varepsilon)\, \bigg\langle \frac{1}{r^4}\bigg\rangle_\varepsilon\,,  \label{68}\\
	\bigg\langle \nabla^k\,\biggl[\frac{\delta^{ij}}{6\,r} - \frac{r^i\,r^j}{2\,r^3}\biggr]_\mathrm{\varepsilon}\,
	\nabla^k \biggl[\frac{\delta^{ij}}{6\,r} - \frac{r^i\,r^j}{2\,r^3}\biggr]_\mathrm{\varepsilon}\bigg\rangle =&\
	\frac{(d-1)(d^2-2d+4)}{4d}\bigg\langle\frac{1}{r^4}\bigg\rangle_\varepsilon 
 =
	\biggl(\frac{7}{6} - \frac{31}{18}\,\varepsilon \biggr)\,\bigg\langle \frac{1}{r^4}\bigg\rangle_\varepsilon + O(\varepsilon)\,,  \label{69}
\end{align}
and use \cite{patkos:17:singlet}
\begin{align}	
	\bigg\langle\frac{1}{r^4}\bigg\rangle_\varepsilon =&\  \bigg\langle\frac{1}{r^4}\bigg\rangle + Z\,\langle\pi\delta^d(r)\rangle\,\biggl(-\frac{2}{\varepsilon} +8\biggr) + O(\varepsilon)\,,
	\label{70}
\end{align}
where $\langle 1/r^4 \rangle$ is defined as integral from a small radius $a$ to infinity with the
$1/a$ and $\ln a+\gamma$ terms subtracted out.
The $1/\varepsilon$ singularity cancels out, and we obtain the final result for the hydrogenic atom as
\begin{align}
	E_\mathrm{hfs2} =&\ E_\mathrm{secA} + E_\mathrm{secB} + E_\mathrm{diam} + E_\mathrm{high}
	\nonumber \\ =&\
	E_{F2}\,	\biggl[\frac{1}{9}\,\bigg\langle V^+\,\frac{1}{(E-H)'}\,V\bigg\rangle
	+\frac{1}{12}\,\bigg\langle V^{+ij}\,\frac{1}{E-H}\,V^{ij}\bigg\rangle
	+ \frac{16}{9}\,\langle \pi\,\delta^3(r) \rangle\,\bigg\langle  \frac{1}{r}\bigg\rangle
	+\frac{5}{12}\,\bigg\langle\frac{1}{r^4}\bigg\rangle
	\nonumber \\ &\
	+ \frac{Z^4}{n^3}\,\frac{10}{3}\biggl( \frac{1223}{540} - \frac{2\,\ln 2}{5} - \frac{3\,\ln 3}{10}  + \ln m\,\alpha\,r_C \biggr)\biggr]\,.  \label{71}
\end{align}
Using the following expectation values with hydrogenic $nS$-state wave functions,
\begin{align}
	\biggl\langle\frac{1}{r^4}\biggr\rangle =&\ \frac{8\,Z^4}{n^3}\,
	\biggl[-\frac{5}{3} + \frac{1}{2\,n} +\frac{1}{6\,n^2}+\gamma+\Psi(n)-\ln\frac{n}{2} +\ln Z \biggr]\,,  \label{72} \\
	\bigg\langle V^+\,\frac{1}{(E-H)'}\,V\bigg\rangle = &\ Z^4\,\biggl(- \frac{56}{3\,n^5} - \frac{24}{n^4} + \frac{8}{3\,n^3}\biggr)\,,  \label{73}\\
	\bigg\langle V^{+ij}\,\frac{1}{(E-H)}\,V^{ij}\bigg\rangle = &\ Z^4\,\biggl(\frac{2}{9\,n^5} - \frac{2}{3\,n^3}\biggr)\,,  \label{74}
\end{align}
we obtain for the hydrogenic $nS$ state
\begin{align}
	E_\mathrm{hfs2}(n) =&\  m\,(Z\,\alpha)^6\,\Big(\frac{g\,m}{2\,M}\Big)^2\,I(I+1)\,
	\frac{10}{3\,n^2}\biggl[ \frac{181}{270} - \frac{3}{10\,n} + \frac{1}{12\,n^2} + \frac{3}{10} \,\ln\frac{4}{3} - \ln(n)  + \gamma + \Psi(n) + \ln m\,Z\,\alpha\,r_C \biggr]\,.  \label{75}
\end{align}
Specifically, for $1S$ state it reduces to
\begin{align}
	E_\mathrm{hfs2}(1) =&\  m\,(Z\,\alpha)^6\,\Big(\frac{g\,m}{2\,M}\Big)^2\,I(I+1)\, \frac{10}{3}\,(0.540\,008 + \ln m\,Z\,\alpha\,r_C)
	\,,\label{76}
\end{align}
which is in perfect agreement with the results of our numerical calculations described in Sec.~II.
Namely, the analytical constant term $0.540\,008$ in Eq.~(\ref{75})
agrees with the result obtained by fitting the all-order numerical data, $2.700(2)/5 = 0.540\,0\,(4)$.

The final result for the helium atom takes the form (in atomic units) 
\begin{align}
E_\mathrm{hfs2} =&\
E_{F2}\,\biggl\{ \sum_a
	\biggl[ \frac{5}{12}\,\bigg\langle\frac{1}{r_a^4}\bigg\rangle
+\frac{10}{3}\biggl( \frac{1223}{540} - \frac{2\,\ln 2}{5} - \frac{3\,\ln 3}{10}  + \ln m\,\alpha\,r_C \biggr)\,Z\,\pi\,\langle\delta^{3}(r_a)\rangle\biggr]
	\nonumber \\ &\
+\sum_{a,b}
\biggl[\frac{1}{27}\,\bigg\langle \vec\sigma_a\,V^+_a\,\frac{1}{(E-H)'}\,\vec\sigma_b\,V_b\bigg\rangle
+\frac{1}{36}\,\bigg\langle \vec\sigma_a\, V_a^{+ij}\,\frac{1}{E-H}\,\vec\sigma_b\,V_b^{ij}\bigg\rangle	
+ \frac{1}{3}\, \bigg \langle \frac{\vec r_a}{r_a^3}\times \vec p_a\,\frac{1}{E-H}\,\frac{\vec r_b}{r_b^3}\times \vec p_a \bigg \rangle\biggr]
\nonumber \\ &\	
+\frac{16}{27}\,\sum_{a,b} \vec\sigma_a\cdot\vec\sigma_b\,
	\langle \pi\,\delta^3(r_a) \rangle\,\bigg\langle  \frac{1}{r_b}\bigg\rangle
-\frac{16}{27}\,\sum_{a \neq b} \vec\sigma_a\cdot\vec\sigma_b\, \bigg\langle \frac{\pi\,\delta^3(r_a)}{r_b}\bigg\rangle
+\frac{1}{18}\,\sum_{a \neq b}\,\vec\sigma_a\cdot\vec\sigma_b\,
	\bigg\langle \frac{1}{r_a^3\,r_b} -3\frac{(\vec r_a\cdot\vec r_b)^2}{r_a^5\,r_b^3}\bigg\rangle \biggr\}
\,,\label{77}
\end{align}
\end{widetext}
where $E_{F2}$ is defined in Eq.~(\ref{EF2}), the operator $V_a$ is defined in Eq.~(\ref{39}),
and $V^{ij}_a$ is defined in Eq.~(\ref{46}).

\section{Numerical results}

We first consider the hydrogenlike atoms.
Using Eq.~(\ref{75}), we obtain the second-order HFS correction to the
$1S$\,--\,$2S$ transition to be $0.86$ kHz for H, $0.05$ kHz for D, and $6.84$ kHz for He$^+$.
Consequently, the contribution for the H\,--\,D isotope shift of the $1S$\,--\,$2S$ transition energy is $0.81$ kHz,
which is twice the previously assumed uncertainty \cite{pachucki:18}.
We, however, do not update the charge radius difference
determined from the H\,--\,D isotope shift, as the corresponding inelastic three-photon exchange
contribution must also be considered, which could be as large as the elastic one calculated here.
Notably, the inelastic  two-photon exchange contribution has already been accounted
for the H\,--\,D  isotope shift in Refs.~\cite{friar:a, friar:b,pachucki:18}.

\begin{table*}
\renewcommand{\arraystretch}{0.92}
\caption{Expectation values of various operators for the $2^1\!S$ and $2^3\!S$ states, in atomic units.}
\label{oprsQ1}
\begin{ruledtabular}
\begin{tabular}{llw{5.8}w{5.8}}
 &                                &   \multicolumn{1}{c}{$2^1S$}          &  \multicolumn{1}{c}{$2^3S$}  \\ \hline
$Q_1 $ & $4 \pi \delta^3 (r_1)$   &   16.455\,169	&	16.592\,071		  \\
$Q_3 $ & $4 \pi \delta^3(r_1)/r_2$&    5.593\,744   &    4.648\,724           \\	
$Q_{53}$ & $1/r_1$				  &    1.135\,408   &    1.154\,664  \\
$Q_{57}$ & $1/r_1^4$   		      &   25.531\,680   & 	25.511\,838		\\
$Q_{65}$ & $[r_1^2 r_2^2 - 3 (\vec{r}_1\cdot\vec{r}_2)^2]/(r_1^5r_2^3)$  &  0.015\,024     & 0.001\,629 \\
\end{tabular}
\end{ruledtabular}
\end{table*}

\begin{table*}
\caption{Second-order matrix elements for $2^1S$ and $2^3S$ states of
He, in atomic units.
\label{table:sec}
}
\begin{ruledtabular}
\begin{tabular}{lllcw{5.8}}
\multicolumn{1}{l}{State} &   \multicolumn{1}{l}{Term} &
    \multicolumn{1}{l}{Expression} &
    \multicolumn{1}{l}{Symmetry} &
            \multicolumn{1}{c}{Value}
\\
\hline\\[-5pt]
&
$S_0$&$ \lbr 2^1S|4\pi [\delta^3(r_1)-\delta^3(r_2)]|2^3S\rbr^2\,[E_0(2^1S)-E_0(2^3S)]^{-1}$
                                                                                        &        &  29\,015.120 \\[1ex]
$2^1S$ &
$S(^3\!S)$&$ \Big<\Big(V_1^+-V_2^+\Big)\,\frac{1}{(E_0-H_0)'}\,\Big(V_1-V_2\Big) \Big>$            & $^3S$ &  27\,703.473\,2     \\
&
$S(^3\!D)$&$ \Big<\Big(V_1^{ij^+}-V_2^{ij^+}\Big)\,\frac{1}{(E_0-H_0)'}\,\Big(V_1^{ij}-V_2^{ij}\Big) \Big>$
                                                                                        & $^3D^e$ &  -7.356\,2 \\
&
$S(^1\!P^e)$&$ \Big\langle\Big(\frac{\vec r_1}{r_1^3}\times\vec p_1+\frac{\vec r_2}{r_2^3}\times\vec p_2\Big)\,
\frac{1}{(E_0-H_0)'}\,
\Big(
\frac{\vec{ r}_1}{r_1^3}\times\vec{ p}_1+\frac{\vec{ r}_2}{r_2^3}\times\vec{ p}_2
\Big)\Big>  $                                    & $^1P^e$  &   -0.012\,469 \\[2ex]
$2^3S$ &
$S(^3\!S)$&$ \Big<\Big(V_1^++V_2^+\Big)\,\frac{1}{(E_0-H_0)'}\,\Big(V_1+V_2\Big) \Big>$            & $^3S$ &      -658.648\,8 \\
&
$S(^1\!S)$&$ \Big<\Big(V_1^+-V_2^+\Big)\,\frac{1}{(E_0-H_0)'}\,\Big(V_1-V_2\Big) \Big>$            & $^1S$  & -29\,425.198\,9 \\
&
$S(^3\!D)$&$ \Big<\Big(V_1^{ij^+}+V_2^{ij^+}\Big)\,\frac{1}{(E_0-H_0)'}\,\Big(V_1^{ij}+V_2^{ij}\Big) \Big>$
                                                                                        & $^3D^e$ &  -7.463\,4 \\
&
$S(^1\!D)$&$ \Big<\Big(V_1^{ij^+}-V_2^{ij^+}\Big)\,\frac{1}{(E_0-H_0)'}\,\Big(V_1^{ij}-V_2^{ij}\Big) \Big>$
                                                                                        & $^1D^e$ &  -7.458\,3 \\
&
$S(^3\!P^e)$&$ \Big\langle\Big(\frac{\vec r_1}{r_1^3}\times\vec p_1+\frac{\vec r_2}{r_2^3}\times\vec p_2\Big)\,
\frac{1}{(E_0-H_0)'}\,
\Big(
\frac{\vec{ r}_1}{r_1^3}\times\vec{ p}_1+\frac{\vec{ r}_2}{r_2^3}\times\vec{ p}_2
\Big)\Big>  $                                    & $^3P^e$    & -0.004\,694 \\
\end{tabular}
\end{ruledtabular}
\end{table*}

Now let us consider the helium atom.
For the singlet states the total electron spin vanishes and, therefore,
$\vec\sigma_1\cdot\vec\sigma_2 = -3$, $(\vec\sigma_1-\vec\sigma_2)^2 =   12$.
Using notations from Tables~\ref{oprsQ1}  and \ref{table:sec}, we write the complete result for the second-order HFS correction
for singlet $S$ states as
\begin{align}	
E_\mathrm{hfs2} =&\	
E_{F2}\,\biggl\{
\frac{1}{9}\,S(^3\!S)+\frac{1}{12}\,S(^3\!D)+ \frac{1}{3}\, S(^1\!P^e)
	\nonumber \\ &\
+\frac{5}{3}\biggl( \frac{1223}{540} - \frac{2\,\ln 2}{5} - \frac{3\,\ln 3}{10}  + \ln m\,\alpha\,r_C \biggr)\,Z\,Q_1
	\nonumber \\ &\
+ \frac{8}{9}\, Q_3 + \frac{5}{6}\,Q_{57} -\frac{1}{3}\, Q_{65} \biggr\}\,.  \label{78}
\end{align}

Our numerical calculations are carried out with the fully correlated exponential basis set
by the method described in Ref.~\cite{yerokhin:21:hereview}. Numerical results
for the first- and second-order matrix elements are summarized in Tables~\ref{oprsQ1}  and \ref{table:sec}.
The contribution
to the ionization energy
is obtained by evaluating Eq.~(\ref{78}) and
subtracting the corresponding result for He$^+(1S)$.
Our final result for the $2^1S$ state of helium is
expressed in the form
\begin{align}
E_\mathrm{hfs2}[\mbox{\rm He}] - E_\mathrm{hfs2}[\mbox{\rm He}^+]=&\	
\frac{E_{F2}}{9}\,S_0\, \Big[1+\delta(2^1S)\Big]\,,
\label{79}
\end{align}
where $S_0$ describes the leading $n=2$ hyperfine mixing and is defined in Table~\ref{table:sec}.
The correction $\delta$
corresponds to the hyperfine mixing with $n\neq 2$ states, 
\begin{align}
\delta(2^1S) =&\ -0.034\,25\,, \label{81}
\end{align}
with the corresponding energy shift $ \delta E_\mathrm{mix} = -2.075$ kHz.

For the triplet states the total electron spin is equal to 1, thus
$\vec\sigma_1\cdot\vec\sigma_2  = 1$,
$(\vec\sigma_1-\vec\sigma_2)^2 = 4$,
$(\vec\sigma_1+\vec\sigma_2)^2 = 8$.
Using notations from Tables~\ref{oprsQ1}  and \ref{table:sec}, the complete result for
the second-order HFS correction for triplet $S$ states can be expressed as
\begin{align}
E_\mathrm{hfs2} =&\
E_{F2}\,\biggl\{
\frac{2}{27}\,S(^3\!S)
+\frac{1}{27}\,S(^1\!S)
+\frac{1}{18}\,S(^3\!D)	
+\frac{1}{36}\, S(^1\!D)
	\nonumber \\ &\hspace*{-4ex}
+\frac{5}{3}\biggl( \frac{1223}{540} - \frac{2\,\ln 2}{5} - \frac{3\,\ln 3}{10}  + \ln m\,\alpha\,r_C \biggr)\,Z\,Q_1
	\nonumber \\ &\hspace*{-4ex} 	
+\frac{1}{3}\,S(^3\!P^e)	
+\frac{5}{6}\,Q_{57}	
+\frac{32}{27}\, Q_1\,  Q_{53}
-\frac{8}{27}\, Q_3
+\frac{1}{9}\, Q_{65} \biggr\}\,. \label{82}
\end{align}
With
numerical results
for the first- and second-order matrix elements from Tables~\ref{oprsQ1}  and \ref{table:sec},
we obtain the contribution
to the ionization energy of the $2^3S$ state as
\begin{align}
E_\mathrm{hfs2}[\mbox{\rm He}] - E_\mathrm{hfs2}[\mbox{\rm He}^+] =&\	
-\frac{E_{F2}}{27}\,S_0\,\Big[1+\delta(2^3S)\Big]\,, \label{83}
\end{align}
where
\begin{align}
\delta(2^3S) =&\ 0.015\,08\,, \label{85}
\end{align}
with the corresponding energy shift $ \delta E_\mathrm{mix} = -0.305$ kHz for the $2^3S$ state.

\begin{table*}
\caption{Breakdown of theoretical contributions to the ${}^3\textrm{He}-{}^4\textrm{He}$ isotope shift of
the $2^1S$--$2^3S$ centroid transition frequencies, for the point nucleus, in kHz.
${\cal E}_\mathrm{mix}$ is the mixing contribution induced by $S_0$.
$\delta {\cal E}_\mathrm{mix}$ comes from $\delta(2^1S)$ and $\delta(2^3S)$ terms in Eqs.(\ref{79}) and (\ref{83}). ${\cal E}_\mathrm{pol}$
is the leading nuclear polarizability contribution at the order $\alpha^5$ (two-photon exchange), see Refs. \cite{pachucki:07:heliumnp, muli:24}.
Physical constants are from Ref.~\cite{mohr:22:codata}.}
\label{tab:is}
\begin{center}
\begin{tabular}{c | w{10.6}w{6.6}w{6.6}w{10.8}w{10.6}}
\hline
\hline\\[-7pt]
Contribution  & \centt{$(m/M)^1$}  & \centt{$(m/M)^2$} & \centt{$(m/M)^3$}   &  \centt{Sum} & \centt{Qi {\em et al.}}\\
\hline\\[-7pt]
 $\alpha^2$          &  -8\,026\,758.512 &     -4\,958.331 & 5.070         & -8\,031\,711.773 & -8\,031\,711.78\\
 $\alpha^4$          &       -2\,496.229 &           2.076 & \textrm{---} &      -2\,494.153 & -2\,494.16\\
 $\alpha^5$          &            56.605 & 0.000(47)    & \textrm{---} &           56.605(47) & 56.59\\
 $\alpha^6$          &             2.732^\mathrm{a}
                            &  \textrm{---}        & \textrm{---} &            2.732 &2.73^\mathrm{b}\\
 $\alpha^6\, {\cal E}_\mathrm{mix}$ &        \textrm{---}&  80.765       & 0.000(70)&        80.765(70) & 80.76\\
 $\alpha^6\, \delta {\cal E}_\mathrm{mix}$  &        \textrm{---}& -1.770       & \textrm{---} &       -1.770 & -1.37\\
 $\alpha^7$          &        -0.210(105) &                &              &             -0.210(105)  & -0.21(11)^\mathrm{b}\\
$\alpha^5\,{\cal E}_\mathrm{pol}$  &             0.198(20)& \textrm{---} & \textrm{---} &           0.198(20) & 0.20(2)^\mathrm{b}\\
\hline\\[-7pt]
Total       &&&&    -8\,034\,067.607(136) & -8\,034\,067.24(19)\\
 \hline \hline
\end{tabular}
\end{center}
$^\mathrm{a}$ Agrees with our former estimate of $2.75\,(69)$ \cite{pachucki:15:jpcrd},\\
$^\mathrm{b}$ Taken from Ref.~\cite{patkos:17:singlet}.
\end{table*}

\section{${2^3S}$-${2^1S}$ isotope shift}

In our previous calculation of the isotope shift of the $2^3S - 2^1S$ transition energy of helium \cite{patkos:17:singlet},
the second-order HFS contribution was accounted for approximately by taking the leading $n=2$ mixing contributions
$S_0$ and neglecting $\delta(2^1S)$ and $\delta(2^3S)$ in Eqs.~(\ref{79}) and (\ref{83}).
It was recently pointed out by Qi and coworkers \cite{qi:24} that these contributions are in fact significant
and shift the theoretical predictions for the isotope shift beyond the previously estimated error bars.
We now extend our previous theory by accounting for the complete second-order HFS correction
calculated in the previous sections.

The individual theoretical contributions to the  ${}^3\textrm{He}-{}^4\textrm{He}$ isotope shift of
the $2^1S$--$2^3S$ centroid energies are summarized  in Table~\ref{tab:is}, in comparison with results of
Qi and coworkers \cite{qi:24}. All numerical values apart from $\delta {\cal E}_{\rm mix}$ are taken from
our previous investigation  \cite{patkos:17:singlet}. Note that the entries $\alpha^6$, $\alpha^7$, and
$\alpha^5 {\cal E}_{\rm pol}$ were not independently reproduced by Qi and coworkers but taken from
our work \cite{patkos:17:singlet}.
The deviation from Qi and coworkers  for $\delta {\cal E}_{\rm mix}$ is due to the fact that they
did not calculate the complete second-order HFS correction but used an approximate representation for it.
Otherwise, we observe very good agreement with their calculation.

The primary source of uncertainty arises from the $\alpha^7$ term, which is estimated basing on the mass scaling of the hydrogenic
$A_{72}$ coefficient. A calculation of the full $\alpha^7$ correction for the helium atom is quite challenging and has not
been attempted so far. Another open challenge is the inelastic second-order HFS contribution. However,
given that the leading (electric dipole) polarizability correction is
only 0.2 kHz, it can be argued to be negligible at the current level of precision.

The theory for the point nucleus summarized in Table~\ref{tab:is} needs to be complemented by
the description of the energy shift arising  due to the nuclear mean square charge radius  $r_C$.
Generalizing the expression for the finite nuclear size (fs) effect for the hydrogenic energy levels
\cite{pachucki:18} to the helium case, we obtain
\begin{align}
E_\mathrm{fs}[{}^A\textrm{He}] =&\
	\frac{2}{3}\, Z\,\alpha^4\,m\,\frac{r^2_C}{\not\!\lambda^2}\,
	\sum_a \langle\,\pi\,\delta^3(r_a)\rangle_M\,
\nonumber \\ \times &\
	\Big\{1
- (Z\,\alpha)^2
\Big[ \ln \Big(Z\,\alpha\,\frac{r_C}{\not\!\lambda}\Big) + \beta\Big]
\nonumber \\ &
+ \alpha\,(Z\,\alpha)\,(4\,\ln 2-5)\Big\}
\equiv C_A\, r^2_C
\,, \label{86}
\end{align}
where $\not\!\!\lambda = 386.159$ fm is the reduced Compton wavelength,
the expectation value of the $\delta$ function is calculated for the case of the finite-mass nucleus, and the nonlogarithmic
relativistic correction $\beta = - 0.41$ is obtained from Ref.~\cite{pachucki:18}, assuming the $1S$ state and the dipole charge distribution
model. The last term in the brackets of Eq.~(\ref{86}) is the radiative fs correction \cite{mohr:22:codata}.

It should be pointed out that, because of the nuclear mass dependence, the fs effect to the isotope shift is not exactly
proportional to the difference of the squared radii (as was assumed in Ref.~\cite{qi:24}) but has a small contribution
beyond this. We therefore write the fs contribution to the  ${}^3\textrm{He}-{}^4\textrm{He}$ isotope shift as
\begin{align}
E_\mathrm{fs}[{}^3\textrm{He}\!-\!{}^4\textrm{He}] =&\ C_3\,r^2_3 - C_4\,r^2_4
\nonumber \\ =&\
\frac{C_3+C_4}{2}\,\big[r^2_3 - r^2_4\big]
+ \frac{C_3-C_4}{2}\,\big[r^2_3 + r^2_4\big]
\nonumber \\ \equiv&\
C\,\big[r^2_3 - r^2_4\big] + D\,\big[r^2_3 + r^2_4\big]\,, \label{87}
\end{align}
where $r_A \equiv r_C[^A\textrm{He}]$ and the last line is the definition of the coefficients $C$ and $D$.  Their numerical values
are given in Table~\ref{tab:tot} .

We are now in the position to determine the mean square charge radius difference of the helium isotopes,
$\delta r^2 = r^2[^3\mbox{\rm He}] - r^2[^4\mbox{\rm He}]$ by comparing the theory prediction with the measured transition frequencies.
Table~\ref{tab:tot} summarizes all experimental and theoretical input required for the determination.
Our result for the mean square charge radius difference of $\delta r^2 =  1.067\,8\,(7)$ fm$^2$
agrees within $1.3\,\sigma$ with the muonic value of $1.063\,6\,(31)$ fm$^2$ \cite{schuhmann:23}.

\section{Summary}

We have derived the complete elastic second-order hyperfine-interaction correction
of order $\alpha^6$ to the centroid energy levels
of hydrogen-like and helium-like ions. The resulting formulas have been
verified through comparisons with our all-order numerical calculations for hydrogen-like ions.
Numerical calculations have been performed for the $2^1S$ and $2^3S$ states of atomic helium,
extending the previous approximate treatment of this effect by Qi {\em et al.} \cite{qi:24}.

The obtained results impact the determination of the nuclear charge radii difference between
$^3$He and $^4$He, largely resolving the previously reported disagreement
between the muonic and electronic helium determinations \cite{werf:23,schuhmann:23}.
From the measured isotope shift of the $2^1S$\,--\,$2^3S$ transition energy
we obtain the mean square charge radius difference of $^3$He and $^4$He,
which is only $1.3\,\sigma$ away from the muonic value.
Given the difficulty of estimating the nuclear polarizability uncertainties in muonic atoms,
we consider this to be good agreement.

Remarkably, the determination of $\delta r^2$ from
the electronic helium spectroscopy is four times more accurate than that from the muonic
helium. This is the consequence of the fact that the  inelastic nuclear effects for muonic atoms 
are much more significant than in the electronic ones.
The observed agreement between electronic and muonic measurements supports the reliability
of nuclear charge radius determinations from muonic atoms, suggesting that they could
be successfully extended to heavier nuclei, which is the objective of  the QUARTET collaboration \cite{nuc_radii}.

\begin{table*}
\caption{Determination of the nuclear charge difference $\delta r^2$ from the measurement of $^3$He\,--\,$^4$He isotope shift in the $2^1S$\,--\,$2^3S$ transition,
in kHz unless specified otherwise. Physical constants are from Ref. \cite{mohr:22:codata}.}
\label{tab:tot}
\begin{ruledtabular}
  \begin{tabular}{l.l}
$E(^3{\rm He},2^1S^{F=1/2} - 2^3S^{F=3/2})$ &  192\,504\,914\,418x.96(17) &Exp. \cite{werf:23}\\
$-E(^4{\rm He},2^1S - 2^3S)$ & -192\,510\,702\,148x.72(20) & Exp. \cite{rengelink:18} \\
$\delta E_{\rm hfs}(2^3S^{3/2})$& -2\,246\,567x.059(5) & Exp. \cite{schluesser:69,rosner:70}\\
$-\delta E_{\rm iso}(2^1S - 2^3S)$ (point nucleus) &8\,034\,067x.607\,(136) & Theory, Table~\ref{tab:is} \\ [1ex]
Sum               &-229x.21(14) & \\
$C$                            &-214x.758\,\,\, {\rm kHz/fm}^2  & this work  \\
$D$                            &0x.017\,\,\, {\rm kHz/fm}^2  & this work  \\[1ex]
$\delta r^2 = r^2(^3\mbox{\rm He}) - r^2(^4\mbox{\rm He})$                   & 1x.067\,8(7)\;{\rm fm}^2             & this work\\
& 1x.069\,3(15)\;{\rm fm}^2 & Ref. \cite{qi:24} \\
& 1x.075\,7(15)\;{\rm fm}^2 & previous \cite{patkos:17:singlet, pachucki:17:heSummary}\\
& 1x.063\,6(6)_\mathrm{exp}(30)_\mathrm{theo}\;{\rm fm}^2 & $\mu^{3,4}$He$^+$ Lamb shift \cite{schuhmann:23} \\
  \end{tabular}
\end{ruledtabular}
\end{table*}
\acknowledgments
We gratefuly acknowledge interesting discussions with F. Hagelstein, A. Antognini, R. Pohl, K. Eikema,  and F. Merkt.
\appendix
\section{Dimensional regularization}\label{app:dimreg}
This appendix is based on Ref.~\cite{patkos:17:singlet}.
In order to extend spin-$ \frac{1}{2} $ into $ d $ dimensions, we define antisymmetric tensor
\begin{equation}
\sigma^{ij} = \frac{i}{2}[\gamma^i,\gamma^j]\, , \label{A01}
\end{equation}
which in the three-dimensional limit simplifies to
\begin{equation}
\sigma^{ij}\overset{d\rightarrow 3}{=} 2\,\epsilon^{ijk}s^k. \label{A02}
\end{equation}
Since we consider only corrections to energy and neglect the hyperfine splitting,
we may perform the angular average
\begin{align}
\sigma_a^{ij}\,\sigma_b^{kl} \rightarrow&\ \sigma_a^{ij}\,\sigma_b^{ij}\,\frac{\delta^{ik}\,\delta^{jl} -\delta^{il}\,\delta^{jk}}{d\,(d-1)}\,
\,,
\label{A03} \\
\sigma_a^{ij}\,\sigma_b^{il} \rightarrow&\ \sigma_a^{ij}\,\sigma_b^{ij}\,\frac{\delta^{jl}}{d} \label{A04}
\,.
\end{align}
For $a=b$,
\begin{align}
\sigma^{ij}\,\sigma^{ij} = d\,(d-1)\,. \label{A05}
\end{align}
We will not always have spin 1/2 for the nucleus.
In the dimensional regularization, we will pretend that $I=1/2$ and use $\sigma_N^{ij}$.
However, instead of the above identity, we will consistently use the notation
\begin{align}
\sigma_N^{ij}\sigma_N^{ij} \equiv 8\,\lbr\vec{I}^{\,2}\rbr_\varepsilon\,, \label{A06}
\end{align}
and drop the subscript $\varepsilon$ once all $1/\varepsilon$ terms are canceled out.

Throughout our calculations, we extensively used the following result for the general $ d
$-dimensional integral,
\begin{align}\
\Omega_d =&\ \frac{2\,\pi^{d/2}}{\Gamma(d/2)}\,, \label{A07}\\
\int \frac{d^dp}{(2\pi)^d} \frac{4\pi}{p^n} e^{i\vec{p}\cdot\vec{r}} =&\ 2^{2-n} \pi^{1-d/2}\frac{\Gamma\left(\frac{d-n}{2}\right)}{\Gamma\left(\frac{n}{2}\right)}r^{n-d}\,. \label{A08}
\end{align}
Two special cases are of particular importance (with $d = 3-2\,\varepsilon$):
\begin{align}
	\mathcal{V}(r) =&\  \int \frac{d^dp}{(2\pi)^d} \frac{4\pi}{p^2} e^{i \vec{p}\cdot\vec{r}} = \frac{C_1}{r^{1-2\varepsilon}}, \label{A09}\\
	\mathcal{V}^{(2)}(r) =&\  \int \frac{d^dp}{(2\pi)^d} \frac{4\pi}{p^4} e^{i \vec{p}\cdot\vec{r}} = C_2\, r^{1+2\varepsilon}\,, \label{A10}
\end{align}
where
\begin{align}
C_1 =&\ \pi^{\varepsilon-1/2}\Gamma(1/2-\varepsilon)\,,\label{A11}\\
C_2 =&\ \frac{1}{4}\pi^{\varepsilon-1/2}\Gamma(-1/2-\varepsilon)\,. \label{A12}
\end{align}
We define also the associated potentials:
\begin{align}
	\mathcal{V}_\rho =&\ 4\pi \int \frac{d^dp}{(2\pi)^d} \frac{\tilde\rho(p^2)}{p^2} e^{i \vec{p}\cdot\vec{r}}, \label{A13}\\
	\mathcal{V}_\rho^{(2)} =&\ 4\pi \int \frac{d^dp}{(2\pi)^d} \frac{\tilde\rho(p^2)}{p^4} e^{i \vec{p}\cdot\vec{r}}, \label{A14}
\end{align}
and in $d=3$
\begin{align}
{V}_\rho =&\ 4\pi \int \frac{d^3p}{(2\pi)^3} \frac{\tilde\rho(p^2)}{p^2} e^{i \vec{p}\cdot\vec{r}}, \label{A15} \\
{V}_\rho^{(2)} =&\ 4\pi \int \frac{d^3p}{(2\pi)^3} \frac{\tilde\rho(p^2)\,e^{i \vec{p}\cdot\vec{r}}-1}{p^4} . \label{A16}
\end{align}
Asymptotic forms of these potentials at $ r\rightarrow\infty $ are:
\begin{eqnarray}
	\mathcal{V}_\rho &\rightarrow& \mathcal{V} + \text{local terms}, \label{A17}\\
	\mathcal{V}_\rho^{(2)} &\rightarrow& \mathcal{V}^{(2)} +  \tilde\rho'(0)\mathcal{V} + \text{local terms}\,, \label{A18}
\end{eqnarray}
and in $d=3$
\begin{eqnarray}
	V_\rho &\rightarrow& \frac{1}{r} + \text{local terms}, \label{A19}\\
	V_\rho^{(2)} &\rightarrow& -\frac{r}{2} + \frac{\tilde\rho'(0)}{r}+\text{local terms}. \label{A20}
\end{eqnarray}

\end{document}